\documentclass[twocolumn,twoside,slac_two]{revtex4}
\usepackage{graphicx}
\usepackage{fancyhdr}
\begin{document}

\title{On the Magnetic Field in Quasar and FR II Large-Scale Jets}

\author{\L . Stawarz}
\affiliation{Harvard-Smithsonian Center for Astrophysics, Cambridge MA, USA}

\author{J. Kataoka}
\affiliation{Tokyo Institute of Technology, Meguro, Tokyo, Japan} 

\begin{abstract}
Here we report systematic comparison of the spectral properties of large-scale jets, hotspots and extended lobes in quasars and FR II radio galaxies recently observed with {\it Chandra} and {\it ASCA}. We argue that if the strong X-ray emission of the jet knots in these objects results from comptonisation of the CMB photons, as usually considered, the powerful large-scale jets are most likely far from the minimum-power condition in the sense that the magnetic field thereby is below equipartition. We also show that the X-ray emission of the hot-spots and lobes in the compiled dataset agrees with the minimum-power condition. In this context, we point out the need for substantial amplification of the magnetic field within the terminal shocks of powerful large-scale jets of quasars and FR II sources.
\end{abstract}

\maketitle

\thispagestyle{fancy}

\section{INTRODUCTION}

The purpose of our study (Kataoka \& Stawarz 2005) is to obtain a rough, but unified picture which may link properties of the large-scale jet-knots, hotspots and radio lobes observed at radio and X-ray frequencies. We apply simple but \emph{uniform} formalism to model broad-band spectra for a large number of sources in terms of synchrotron and inverse-Compton processes (rather than to model individual sources in a sufficiently detailed manner). Our data analysis was based on a sample consisting of 26 radio galaxies, 14 quasars, and 4 blazars, and included 56 jet-knots, 24 hotspots, and 18 radio lobes. For all of these objects we collected the existing data at well sampled radio (5 GHz) and X-ray (1 keV) frequencies. Details of the model are given in~\cite{kat05}.

In the analysis, we applied a simple formulation of computing an equipartition magnetic field strength $B_{\rm eq}$ from an observed radio flux measured at a radio frequency 5 GHz. Next, we calculated the ``expected'' synchrotron self-Compton and external CMB-Compton luminosities for $B_{\rm eq}$, to compare them with the observed 1 keV luminosities. Taking the results obtained into account, and analyzing additionally the observed broad-band spectral properties of the compiled sources (including optical fluxes), we followed the ``conservative'' classification of the discussed X-ray sources into three groups, namely
\begin{itemize}
\item synchrotron involving single/broken power-law electron energy distribution (SYN),
\item synchrotron self-Compton (SSC), and
\item external Compton of CMB photons (EC).
\end{itemize}
Here we present some aspects of our study regarding large-scale structures in radio loud quasars and FR II radio galaxies. For the alternative, synchrotron interpretation of the X-ray jet-knots in powerful radio sources see~\cite{sta04}.

\begin{table}
\begin{center}
Table 1. Source classification of jets, hotspots, and lobes.

\vspace{3mm}

\begin{tabular}{lccc}
\tableline\tableline
      & Jet-knot & Hotspot & Lobe \\
\tableline
QSO(CD) & 19  & 2 &  0 \\
QSO(LD)  & 7   & 9  & 6  \\
RG(FR I) & 22    &0  & 3 \\
RG(FR II) & 1    &13  & 9 \\
BLZR & 7    &0  & 0 \\
\tableline
SYN & 25  & 7 &  0 \\
SSC  & 4   & 16  & 1  \\
EC & 27    &1  & 17 \\
\tableline
\end{tabular}
\end{center}
\end{table}

\section{RESULTS}

A number of quasar jet-knots classified as the SSC or EC sources (see table 1) seem extremely bright in X-ray. This inevitably causes a large discrepancy between the ``expected'' (for sub-relativistic jet velocities) and ``observed'' X-ray fluxes, a fact that is well known from the previous studies reported in the literature (see, e.g.,~\cite{staw04} for a review). There are two formal possibilities \emph{in a framework of homogeneous one-zone EC emission region model} to account for the discussed discrepancy:
\begin{itemize} 
\item the equipartition hypothesis may not be valid for the jet-knots considered, or
\item relativistic beaming effects are significant enough to allow the minimum-power condition to be met.
\end{itemize}

\begin{figure}
\includegraphics[width=65mm,angle=-270]{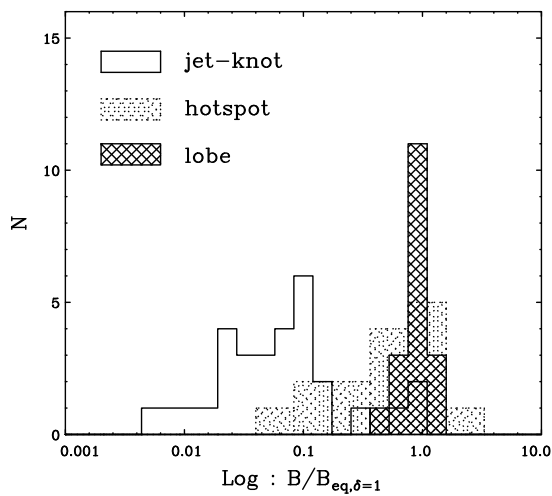}
\caption{Distribution of the ratio between the magnetic field  $B$ (for $\delta$ = 1) and the equipartition value $B_{\rm eq, \, \delta = 1}$.}
\label{f1}
\end{figure}

Figure 1 shows the ratio of the magnetic field $B$ estimated for the discussed objects (for $\delta = 1$) to the equipartition value $B_{\rm eq, \, \delta=1}$. Interestingly, $B$ values for the lobes and for most of the hotspots are almost consistent with the equipartition ($B$/$B_{\rm eq,\delta=1}$ $\sim$ 1), whereas those for the non-SYN jet-knots and for some of the hotspots are much weaker than expected ($B$/$B_{\rm eq, \delta=1}$ $\sim$ 0.01$-$0.1).

As an alternative, we consider a case when the difference between the ``expected'' and ``observed'' X-ray fluxes is due to the relativistic beaming effect, and the minimum-power condition is fulfilled~\cite{tav00,cel01,har02}. The Doppler factors thus calculated are shown in figure 2. One can see that the lobes and hotspots exhibit relatively narrow distribution at $\delta$ $\sim$ 1, whereas for most of the jet-knots large beaming factors of $\sim$ 10 are required, as noted before by many authors.

\subsection{Strongly Beamed or Far-From-Equipartition?}

Usually, in applying the EC model to the quasar jet-knots' X-ray emission, the idea of a sub-equipartition magnetic field is rejected since it implies a very high kinetic power of the jets. For this reason, large values for the jet Doppler factors are invoked. However, it is well known that the {\it VLA} studies of the large-scale jets in quasars and FR IIs reported by~\cite{war97} indicate that bulk Lorentz factors of the radio-emitting plasma in these sources cannot be much greater than $\Gamma$ $\sim$ 3. The discrepancy between this result and the requirement of the minimum-power EC model for $\Gamma$ $>$ 10 is typically ascribed to the jet radial velocity structure, namely that the radio emission originates within the slower-moving jet boundary layer and the inverse-Compton X-ray radiation is produced within the fast jet spine~\cite{ghi01}. 

\begin{figure}
\includegraphics[width=65mm,angle=-270]{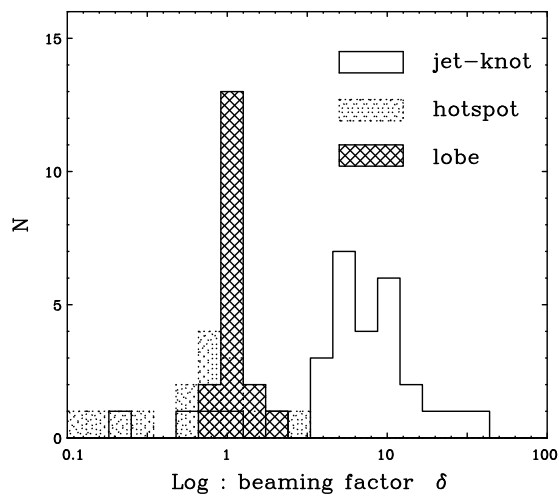}
\caption{Distribution of the required beaming factor $\delta$ for $B$ = $B_{\rm eq}$.}
\label{f2}
\end{figure}

While it is true that jet radial stratification can indeed significantly influence the jet-counterjet brightness asymmetry ratio, one should be aware that by postulating different sites for the origin of radio and X-ray photons, \emph{homogeneous one-zone models for the broad-band knots' emission can no longer be preserved}. In particular, in such a case one has to specify exactly what fraction of the jet radio emission is produced within the spine and what fraction within the boundary layer, what exactly the jet velocity radial profile is, and what the magnetic field strength is in each jet component, etc. Without such a discussion one cannot simply use the observed radio flux of the jet to construct the broad-band spectral energy distribution of the knot region, i.e. simply estimate the expected inverse-Compton flux by means of equipartition magnetic field derived from the radio observations. If one insists on applying the homogeneous one-zone model (as a zero-order approximation), self-consistency requires a consideration of $\Gamma$ $\lesssim$ 5. In such a case, a departure from the minimum power conditions within the non-SYN X-ray jets is $inevitable$ and hence powerful jets are most likely $particle$ $dominated$. The jet magnetic field must be then significantly amplified in the hotspot, where an approximate equipartition is expected to be reached.

\subsection{Magnetic Field Strength}

\begin{figure}
\includegraphics[width=65mm,angle=-270]{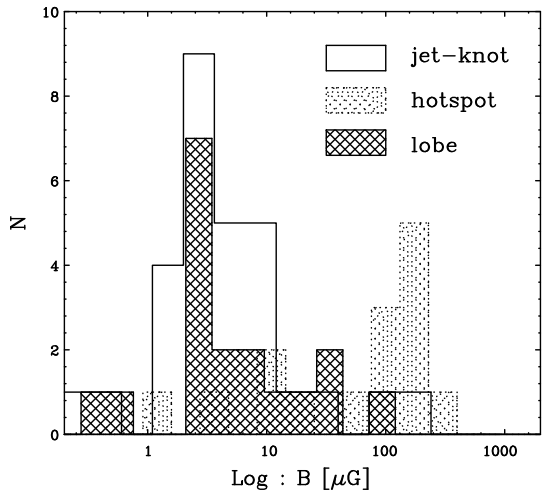}
\caption{Distribution of the evaluated magnetic field, $B$, for the case of $no$ relativistic beaming ($\delta$=1).}
\label{f3}
\end{figure}

Figure 3 shows the distribution of the ``best-fit'' magnetic field $B$ if we allow for the deviation from the equipartition condition and assume nonrelativistic velocities for the emitting regions (which, in the case of the jet-knots, is rather only a formal hypothesis). One finds that both the non-SYN jet-knots and radio lobes are distributed around $B$$\sim$ 1$-$10$\mu$G, whereas hotspots have a relatively narrow peak at higher field strength, $B$ $\sim$ 50$-$300$\mu$G, plus a ``tail'' extending down to $\sim$$\mu$G.

Figure 4 shows the distribution of an equipartition magnetic field in the framework of relativistically moving jet model. Similarly to figure 3, we find again that the narrowly distributed strength of the magnetic field in the hotspots, $B_{\rm eq}$ $\sim$ 100$-$500$ \mu$G, is an order of magnitude larger than that of the jet-knots and radio lobes.

\subsection{Is the EC Hypothesis Correct?}

We have discussed two different versions of the EC model to account for extremely bright X-ray jet-knots: (1) the non-equipartition case and (2) the significant relativistic beaming case. Both are in many ways problematic. Our next concern is to attempt to prove in general the postulated inverse-Compton origin of the X-ray photons. One possibility for doing it is to look for $L_{\rm X}$/$L_{\rm R}$ $\propto$ $(1+z)^4$ behavior within a large sample of EC sources~\cite{sch02,che04}. 

Figure 5 shows the distribution of the flux ratio $L_{\rm 1keV}$/$L_{\rm 5GHz}$ as a function of $z$ for the compiled dataset. The dotted line shows $\propto$ $(1+z)^4$ relation which fits the highest $z$ data point (GB~1508+5714; $z$ = 4.3) just to help guide the eyes. Although the data sample is still poor, we may say that no clear trend can be seen in this plot. Furthermore, we notice that the $L_{\rm X}$/$L_{\rm R}$ ratio is widely distributed even in the same objects. Such a difference is not easy to explain in the framework of model (1), since we have to assume an order of magnitude increase in the magnetic fields along the jet. In the framework of the relativistic beaming hypothesis (2) one may possibly explain such variation by postulating the decrease of the bulk Lorentz factor along the flow and only moderate changes in magnetic field. In this case, however, one has to explain what causes significant deceleration of the jet, which preserves its excellent collimation, with no significant radiative energy losses.

\begin{figure}
\includegraphics[width=65mm,angle=-270]{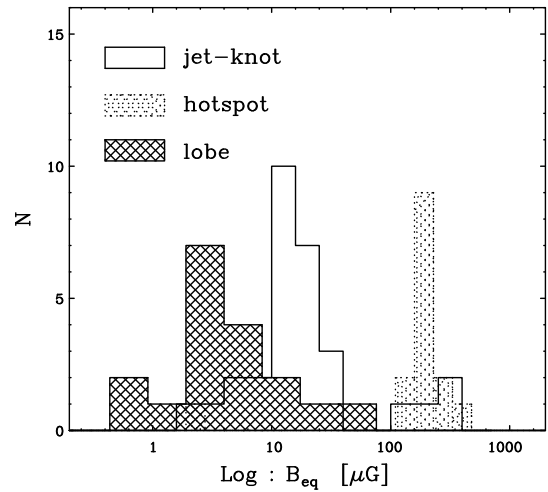}
\caption{Equipartition magnetic field for relativistically moving jet model.}
\label{f4}
\end{figure}

Figure 6 shows the Doppler beaming factor $\delta$ required in the EC model to obtain $B$ $=$ $B_{\rm eq}$, versus the redshifts $z$ of the jet-knots classified here as the EC ones. There are two possible explanations for the noted $\delta$--$z$ anticorrelation. If reflecting physical property, it would mean that the distant large-scale quasar jets are less relativistic than their nearby analogues but similarly close to the equipartition, $or$ that both low- and high-$z$ quasar jets are only mildly relativistic on large scales but closer to the minimum-power condition when located at large redshifts. Neither option appears to be particularly natural, especially as the high-$z$ quasar cores seem to be comparable to their low-$z$ counterparts~\cite{bas04}. On the other hand, differences in velocity and energy content of the large-scale jets may not reflect differences in the central engines, but rather differences in the surrounding galactic or intergalactic medium. The second possibility for understanding $\delta$--$z$ anticorrelation is however that it is simply an artifact of the applied but inappropriate EC model.

\section{DISCUSSION}

Below, we point out three issues emerging from our study in relation to large-scale jets in quasar and FR II radio galaxies:

\begin{figure}
\includegraphics[width=65mm,angle=-270]{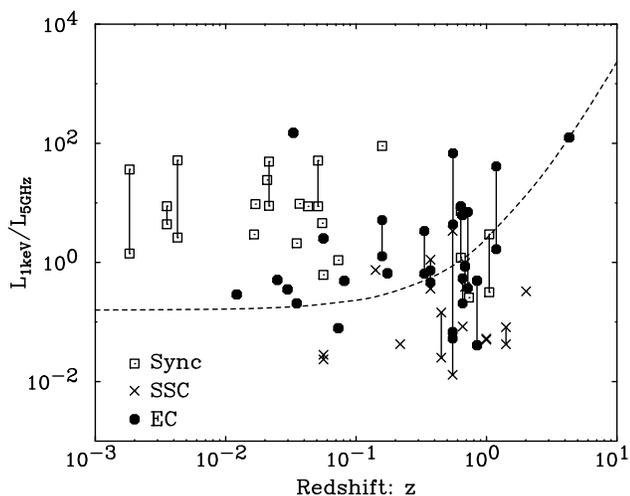}
\caption{Luminosity ratio $L_{\rm 1keV}$/$L_{\rm 5GHz}$ as a function of redshift for SYN, SSC and EC soures.}
\label{f5}
\end{figure}

{\bf (1)} If the strong X-ray emission of the jet knots in these objects results from comptonisation of the CMB photons, as usually considered, the powerful large-scale jets are most likely far from the minimum-power condition in the sense that the magnetic field thereby is below equipartition. The jet magnetic field must be then significantly amplified in the hotspot, where an approximate equipartition is expected to be reached. However, few theoretical investigations of this issue has been reported~\cite{dey02}.

{\bf (2)} Pressure of radio-emitting electrons within the lobes of quasars and FR IIs computed from the equipartition condition --- justified by the presented analysis --- is often found to be below the thermal pressure of the ambient medium, which challenges the standard model for the evolution of powerful radio sources. Such a discrepancy can however be removed by postulating the presence of non-radiating relativistic protons within the lobes. As proposed by~\cite{ost01}, viscous acceleration of cosmic rays taking place at the turbulent boundary layers of relativistic jets can provide an energetically important population of such high energy protons escaping from the jet to the lobes along all of its length. Interestingly, this would imply total energy outputs of powerful jets systematically $larger$ than that implied by analysis of the lobes' radio emission alone. This would be consistent with deviation from the minimum-power condition within the considered jets themselves, as discussed here.

{\bf (3)} Recent analysis by~\cite{sta05} indicates that the magnetic field in M~87 kpc-scale jet cannot be smaller than the equipartition value referring solely to radiating electrons. Such a strong magnetic field can be most likely ascribed to turbulent dynamo processes connected with the entrainment processes important in controlling dynamics (deceleration) of low-power FR I jets (see a discussion in~\cite{sta05}). It is therefore tempting to speculate that in the case of powerful quasar and FR II large-scale jets the subequipartition magnetic field is consistent with the entrainment processes being much less effective for these objects than in the case of their low-power analogues.

\begin{figure}
\includegraphics[width=65mm,angle=-270]{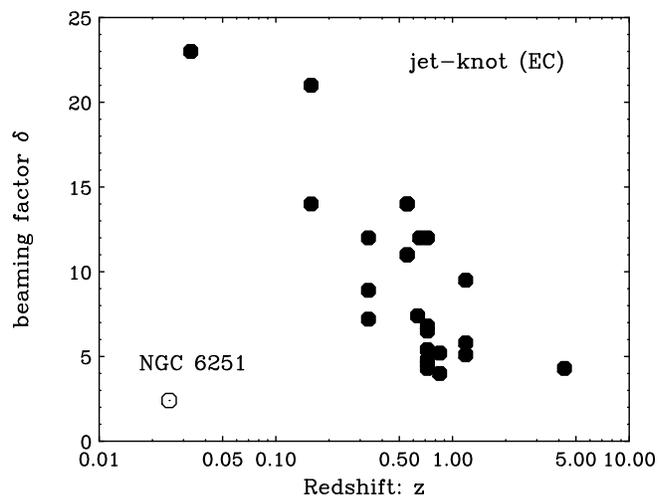}
\caption{Expected beaming factor $\delta_{\rm EC}$ for $B$ = $B_{\rm eq}$, as a function of redshift for EC jet-knot sources. $Open$ $circle$ shows exceptional FR~I radio galaxy NGC~6251.}
\label{f6}
\end{figure}

\bigskip
\begin{acknowledgments}
We would like to thank Fumio Takahara and Micha\l \, Ostrowski for fruitful discussion and constructive comments. J.K. acknowledges a support by JSPS.KAKENHI (14340061). \L .S. was supported by the grant PBZ-KBN-054/P03/2001 and partly by the Chandra grants G02-3148A and G0-09280.01-A.
\end{acknowledgments}


\bigskip 
\begin{thebibliography}{99} 

\bibitem{bas04}
Bassett, L.C., Brandt, W.N., Schneider, D.P., Vignali, C., Chartas, G., \& Garmire, G.P. 2004, AJ, 128, 523
\bibitem{cel01}
Celotti, A., Ghisellini, G., \& Chiaberge, M. 2001, MNRAS, 321, L1
\bibitem{che04}
Cheung., C.C. 2004, ApJ, 600, L23
\bibitem{dey02}
De Young, D. S. 2002, NewAR, 2002, 46, 393
\bibitem{ghi01}
Ghisellini, G., \& Celotti, A. 2001, MNRAS, 327, 739
\bibitem{har02}
Harris, D.E., \& Krawczynski, H. 2002, ApJ, 565, 244
\bibitem{kat05}
Kataoka, J., \& Stawarz, \L . 2005, ApJ, in press (astro-ph/0411042)
\bibitem{ost01}
Ostrowski, M., \& Sikora, M. 2001, in Proc. 20th Texas Symposium on Relativistic Astrophysics, eds. J.C. Wheeler \& H. Martel (Austin), AIP Conf. Ser. 586, 865
\bibitem{sch02}
Schwartz, D.A. 2002, ApJ, 569, L23
\bibitem{staw04}
Stawarz, \L . 2004, ChJAA, in press (astro-ph/0310795)
\bibitem{sta04}
Stawarz, \L ., Sikora, M., Ostrowski, M., \& Begelman, M.C. 2004, ApJ, 608, 95
\bibitem{sta05} 
Stawarz, \L ., Siemiginowska, A., Ostrowski, M., \& Sikora, M. 2005, ApJ, submitted (astro-ph/0501597)
\bibitem{tav00}
Tavecchio, F., Maraschi, L., Sambruna, R.M., \& Urry, C.M. 2000, ApJ, 544, L23
\bibitem{war97}
Wardle, J.F.C., \& Aaron, S.E. 1997, MNRAS, 286, 425

\end{thebibliography}
\end{document}